\documentclass[a4paper, aps, tightenlines, nofootinbib, 12pt]{revtex4}
\usepackage{hyperref}
\usepackage{amsmath}
\usepackage{bbm}
\newsavebox{\myhbar}
\savebox{\myhbar}{$\hbar$}
\usepackage{graphicx}
\usepackage{enumerate}
\usepackage{mdframed}
\usepackage[export]{adjustbox}
\usepackage{braket}
\usepackage{url}
\usepackage{wasysym}

\DeclareMathOperator{\arctanh}{arctanh}
\DeclareMathOperator{\sech}{sech}
\DeclareMathOperator{\Tr}{Tr}
\usepackage{subfig}
\usepackage{graphicx}
\graphicspath{ {./images/} }
\usepackage{float}
\renewcommand*{\hbar}{\mathalpha{\usebox{\myhbar}}}
\begin{document}
	
	\title{Quantum Stirling heat engine with squeezed thermal reservoir} \author {  Nikolaos Papadatos}
	\email{n.papadatos@upatras.gr}
	\affiliation{ Department of Physics, University of Patras, 26500 Greece}
	
	\begin{abstract}
	We analyze the performance of a quantum Stirling heat engine (QSHE), using a two level system and the harmonic oscillator as the working medium, that contacts with a squeezed thermal reservoir and a cold reservoir. First, we derive closed-form expressions for the produced work and efficiency which strongly depends on the squeezing parameter $r_h$. Then, we prove that the effect of squeezing heats the working medium to a higher effective temperature which leads to better overall performance. In particular, the efficiency increases with the degree of squeezing surpassing the standard Carnot limit, when the ratio of temperatures of hot and cold reservoir is small. Furthermore, we derive the analytical expressions for the efficiency at maximum work and the maximum produced work in the high and low temperature regime and we find that at extreme temperatures the squeezing parameter $r_h$ does not affect the performance of the QSHE. Finally, the performance of the QSHE depends on the nature of the working medium. 
	\end{abstract}
	
	\maketitle
	

	\section{Introduction}
		\subsection{Motivation}

Over the past few years, the emerging field of quantum thermodynamics has attracted an increasing attention to the scientific community, trying to find the quantum origin of the laws of thermodynamics \cite{Anders}. Fundamental thermodynamic variables such as temperature $T$, internal energy $U$, etc., have been investigated extensively in terms of classical, statistical, quantum and relativistic quantum thermodynamics \cite{Bimalendu}-\cite{Naka}. Furthermore, the comprehension of the laws of classical thermodynamics is attainable, in the context of quantum heat engines (QHEs), comparing thermodynamic quantities such as work and efficiency in terms of quantum variables.\\
The QHE was first studied by Scovil and Schulz-DuBois in 1959 using a 3-level maser \cite{Scovil}. Thenceforward, the study of performance of QHEs \cite{Alickiengine} has paved the way to new fields of research. In particular, the  work and efficiency of QHEs \cite{Alicki}, \cite{Eitan Geva and Ronnie Kosloff} have been analyzed with respect to the role of coherence, \cite{Scully1}-\cite{Hammam}, the impact of entanglement \cite{Zhangentangled}-\cite{He} the effect of a many-particle working medium  \cite{Beau}-\cite{Jussiau}, the effect of the degeneracy factor in the energy levels \cite{Pena} \cite{Barrios}, fast Otto-cycle \cite{Erdmanfast}, endoreversible Otto engine with harmonic oscillator as working medium  \cite{Deffnerendore}, endoreversible engines \cite{Deffnerendore}, \cite{Smith}, \cite{Myersndore}, an engine based on many body localization as a working fluid \cite{Halpern}.

A new era of quantum thermodynamics is the relativistic quantum thermodynamics, especially in the context of QHEs. In particular, the construction of a QHE that includes relativistic effects. The following literature is indicative of the need for more research.
The performance of QHEs has been analyzed with a relativistic particle confined in the one-dimensional potential well as working substance \cite{Pritam}, with the thermodynamical properties of relativistic and quantum fields\cite{Bruschi}, with non-relativistic working medium interacting with a relativistic bath \cite{nikos1}, with QHEs based on Unruh effect \cite{UnruhXu}, \cite{UnruhGray}.

One of the goals of studying quantum machines is to find ways to improve its overall performance \cite{Quantum Afterburner}, \cite{Abah0}. Studies have emphasized that Reinforcement-Learning \cite{Erdman} and nonequilibrium quantum reservoirs \cite{Abah1} such as coherent \cite{Quancohere}-\cite{Hardalcohere} and correlated \cite{Dillenschneidercorrel} can improve the performance of a quantum heat engine. Additionally, the concept of squeezed thermal reservoir in quantum thermodynamics has started to create an increasing interest. The quantum Otto heat engine whose working medium is in contact with squeezed reservoirs has been studied extensively and presents some very interesting features. In particular, the efficiency at maximum power increases with the degree of squeezing, surpassing the standard Carnot limit \cite{HuangSQUE}-\cite{Manzano}.
In the real world, the efficiency at maximum power is more important quantity instead of maximum efficiency, at which power is zero \cite{Leff}.  The efficiency at maximum power was first studied by Curzon and Ahlborn (CA) \cite{Curzon} and they found that $\eta_{CA}=1-\sqrt{\frac{T_c}{T_h}}$, where $T_c$ and $T_h$ is the temperature of the cold and hot reservoir respectively.

The question arises as to what impact the squeezed thermal bath will have to another thermodynamic cycle, such as the Stirling thermodynamic cycle that has received less attention. Stirling thermodynamic cycle has two isothermal and two isochoric processes \cite{Quan} and produces more work than that of the Otto cycle under the same parameters or equal efficiency conditions and operates a wider range of coupling strengths \cite{akmak}.


The aim of this paper is to study the performance of a QHE  in which the working medium interacts with a pure cold reservoir and a squeezed thermal reservoir \cite{Scullyquantumpro}-\cite{Dung2}. In particular, we analyze the quantum equivalent of the classical Stirling cycle \cite{Chenstirling}-\cite{Purkait} in which the hot reservoir is replaced by a thermal squeezed reservoir. We observe that the squeezing parameter $r_h$ improves the performance of QSHE, especially in the small ratio between hot and cold reservoir which means that the engine has less energetic cost. On the other hand, the generation of squeezing has a thermodynamic cost which depends on the specific configuration employed \cite{Galve}, \cite{Zagoskin}. Generally speaking, when the cost of generating a squeezed state is included in the computation of efficiency of the engine, the squeezed reservoirs are costly \cite{Singh}. However, we focus on studying the performance of QSHE without including the energy losses from the generation of squeezing.

\subsection{Analysis and results}
Our QHE consists of a working medium that is weakly coupled to pure cold reservoir and a squeezed thermal reservoir on alternate pace according to the Stirling thermodynamic cycle. We analyse two cases as far as the working medium is concerned, a two level system and a harmonic oscillator.
Our results are the following:
\begin{enumerate}[(i)]
\item We derive closed-form expressions for the produced work and efficiency which strongly depends on the squeezing parameter $r_h$. 
 
\item In both cases of working mediums, the produced work and the efficiency of the QSHE increases with the degree of squeezing, see Figs. 1-9.
\item The performance of QSHE, i.e., the efficiency and the total produced work depends on the nature of the working medium, see Figs. 1-9.
\item The effective temperature due to squeezing is always higher than the temperature of the thermal reservoir.
\item We derived the analytic expression for the efficiency at maximum work and the maximum produced work in the high and low temperature regime.
\item In the low temperature regime, the efficiency of the QSHE with a two level system as a working medium coincides with the efficiency of quantum Otto heat engine. 
\item At extreme temperatures the squeezing parameter $r_h$ does not affect the performance of the QSHE.
\end{enumerate}

The structure of this paper is the following. In Sec.\ref{Quantum Stirling heat engine - Two-level atom} we examine the first case of QSHE with a two level system as a working medium.  In Sec.\ref{Quantum Stirling heat engine - HO}, we examine the second case of QSHE with harmonic oscillator as a working medium.  The conclusions are presented in Sec. \ref{Conclusions}.

\section{Quantum Stirling heat engine - Two level atom}
\label{Quantum Stirling heat engine - Two-level atom}

In this section, we use a two level system (TLS) of frequency $\omega$ as the working medium. The Hamiltonian is $H=\frac{1}{2}\hbar\omega \sigma^z$. The TLS interacts with the squeezed thermal reservoir at temperature $T_h$, squeeze parameter $r_h$ and the phase $\phi$.
The density matrix equation is \cite{BrePe07}

\begin{align}
	\frac{\partial \hat{\rho}}{\partial \tau} = & \Gamma [N+1]  \left( \hat{\sigma}_-\hat{\rho}\hat{\sigma}_+  -\frac{1}{2}  \hat{\sigma}_+\hat{\sigma}_- \hat{\rho}  -  \frac{1}{2}  \hat{\rho}  \hat{\sigma}_+\hat{\sigma}_-  \right)\nonumber + \Gamma N  \left(   \hat{\sigma}_+ \hat{\rho}\hat{\sigma}_-  -\frac{1}{2}  \hat{\sigma}_- \hat{\sigma}_+ \hat{\rho}  -  \frac{1}{2}  \hat{\rho} \hat{\sigma}_- \hat{\sigma}_+\right)
\nonumber -\\
&-\Gamma M     \hat{\sigma}_+ \hat{\rho}\hat{\sigma}_+-  \Gamma M^{*}  \hat{\sigma}_-\hat{\rho}  \hat{\sigma}_-,
\end{align}

where

\begin{align}
	\label{N}
	N=n\cosh(2r_h)+\sinh^2(r_h),
\end{align}
  and
 \begin{align}
 	M=-\cosh(r_h)\sinh(r_h)e^{i\phi}(2n+1),
 \end{align}
 and
\begin{align}
	\label{n}
n=\frac{1}{e^{\frac{\hbar\omega}{k_BT}}-1},
\end{align}  
   is the mean number of quanta.
We notice that the asymptotic state does not depend on the squeezed phase $\phi$.
\begin{eqnarray}
		\label{asymptoticTLS}
	\rho = \frac{1}{(2n+1)\cosh(2r) }
	\bordermatrix{&              &              \cr
		& n\cosh(2r)+\sinh^2(r)
		& 0 \cr
		& 0
		& n\cosh(2r)+ \cosh^2(r)\cr},
\end{eqnarray}
equivalently, we have
\begin{eqnarray}
	\label{asymptoticTLS}
	\rho = \frac{1}{(2N+1)}
	\bordermatrix{&              &              \cr
		& N
		& 0 \cr
		& 0
		& N+1\cr}.
\end{eqnarray}
 
The expectation value of energy is given by
\begin{eqnarray}
	\label{energygeneral}
	U=\langle \hat{E} \rangle = \Tr(\hat{\rho} \hat{H}),
\end{eqnarray}

thus,

\begin{eqnarray}
	\label{energyTLSst}
U = - \frac{\hbar\omega}{2 [2N+1]}.
\end{eqnarray}

The quantum Stirling heat engine is the quantum counterpart of classical Stirling thermodynamic cycle. It consists of two isothermal prosesses and two isochoric processes. In the first case, as the working medium, we have a two level system that interacts with a hot squeezed reservoir and a cold bath on alternate pace so as to perform a quantum Stirling thermodynamic cycle. \\
The general quantum Stirling cycle has the following four repeated cyclically steps:

\begin{enumerate}[(i)]
	
	\item Isothermal compression:   $A(\omega_2,T_h)\rightarrow B(\omega_1,T_h)$. Initially, the TLS contacts with the hot squeezed reservoir at a constant temperature $T_h$ that can be described by a unitary squeezed operator with squeezed parameters $r_h$ and $\phi$. During this process the mean number of quanta and the energy gaps must change simultaneously in order for the system to remain in an equilibrium state.
	The heat absorbed by the working medium is $Q_{AB}=T_h\Delta S_{AB}$.
	
	\item Isochoric compression: $B(\omega_1,T_h)\rightarrow C(\omega_1,T_c)$. In the second stroke, the TLS is disconnected from the hot squeezed bath and is connected with the cold bath at temperature $T_c$ with fixed frequency.  
	
	\item  Isothermal expansion:    $C(\omega_1,T_c)\rightarrow D(\omega_2,T_c)$. In the third stroke, the TLS is still coupled with the cold bath and the frequency increases from  $\omega_1\rightarrow\omega_2$. The heat released is $Q_{CD}=T_c\Delta S_{CD}$.
	\item  Isochoric expansion:   $ D(\omega_2,T_c)\rightarrow A(\omega_2,T_h)$. Finally, at constant frequency the TLS is disconnected from the cold bath and is connected  with the hot squeezed bath at temperature $T_h$ with fixed frequency.
\end{enumerate}

The entropy for a TLS at thermal equilibrium is 

\begin{align}
S[\hat{\rho}]=-k_B\Tr(\hat{\rho}\log\hat{\rho}),
\end{align}
thus
\begin{align}
\begin{split}
	S=&-\frac{1}{(2n+1)\cosh 2r_h}k_B\Big((\cosh^2r_h+n\cosh 2r_h)\log\frac{2n+1+(\cosh2r_h)^{-1}}{4n+2}+\\
	+&(n\cosh2r_h+\sinh^2 r_h)\log\frac{n+\sinh^2r_h\sech^22r_h}{2n+1}\Big).
\end{split}
\end{align}


  We compute the entropy change in process $AB$
\begin{align}
\label{entropyTLS}
\begin{split}
     &S_{AB}=S(B)-S(A)=\\
 =&k_B\log\Big[\frac{\cosh\Big(\frac{\hbar\omega_1}{2k_BT_h}\Big)}{\cosh\Big(\frac{\hbar\omega_2}{2k_BT_h}\Big)}\Big]+F_{r_h}(\omega_2)\frac{\hbar\omega_2}{2T_h}\tanh\Big(\frac{\hbar\omega_2}{2k_BT_h}\Big)-F_{r_h}(\omega_1)\frac{\hbar\omega_1}{2T_h}\tanh\Big(\frac{\hbar\omega_1}{2k_BT_h}\Big)+\\
 &+\frac{1}{2}k_B\log\Bigg[\frac{1+(1-S^2_{r_h})\sinh^2\Big(\frac{\hbar\omega_2}{2k_BT_h}\Big)}{1+(1-S^2_{r_h})\sinh^2\Big(\frac{\hbar\omega_1}{2k_BT_h}\Big)}\Bigg],
\end{split}
\end{align}

where
\begin{align}
 F_{r_h}(\omega_i)=S_{r_h}\Bigg\{1+\frac{k_BT_h}{\hbar\omega_i}\log\Bigg[\frac{1+S_{r_h}+e^{-\frac{\hbar\omega_i}{k_BT_h}}(1-S_{r_h})}{1+S_{r_h}+e^{\frac{\hbar\omega_i}{k_BT_h}}(1-S_{r_h})}\Bigg]\Bigg\},
\end{align}
and
\begin{align}
 S_{r_h}=\sech(2r_h).
\end{align}

Consequently, the heat from this process is, 
\begin{align}
	Q_{AB}=T_h\Big[S(B)-S(A)\Big],
\end{align}
from (\ref{entropyTLS}) we obtain
\begin{align}
\begin{split}
\label{QAB}
		Q_{AB}&=k_BT_h\log\Big[\frac{\cosh\Big(\frac{\hbar\omega_1}{2k_BT_h}\Big)}{\cosh\Big(\frac{\hbar\omega_2}{2k_BT_h}\Big)}\Big]+F_{r_h}(\omega_2)\frac{\hbar\omega_2}{2}\tanh\Big(\frac{\hbar\omega_2}{2k_BT_h}\Big)-F_{r_h}(\omega_1)\frac{\hbar\omega_1}{2}\tanh\Big(\frac{\hbar\omega_1}{2k_BT_h}\Big)+\\
		&+\frac{1}{2}k_BT_h\log\Bigg[\frac{1+(1-S^2_{r_h})\sinh^2\Big(\frac{\hbar\omega_2}{2k_BT_h}\Big)}{1+(1-S^2_{r_h})\sinh^2\Big(\frac{\hbar\omega_1}{2k_BT_h}\Big)}\Bigg].
\end{split}
\end{align}
The heat absorbed in the isothermal process AB using a thermal squeezed reservoir (\ref{QAB}), has the usual form but with coefficients that contains the squeezing parameter $r_h$.\\
In the limit $r_h\rightarrow 0$, Eq \ref{QAB} takes the form,\\

\begin{align}
		Q_{AB}=k_BT_h\log\Big[\frac{\cosh\frac{\hbar\omega_1}{2T_h}}{\cosh\frac{\hbar\omega_2}{2k_BT_h}}\Big]+\frac{\hbar\omega_2}{2}\tanh\Big(\frac{\hbar\omega_2}{2k_BT_h}\Big)-\frac{\hbar\omega_1}{2}\tanh\Big(\frac{\hbar\omega_1}{2k_BT_h}\Big)
\end{align}
The work output per cycle according to the first law of thermodynamics \cite{AlickiHistory}, \cite{Kieu} 
\begin{align}
Q_{AB}=\Delta U_{AB}+W_{AB}.	
\end{align}


The expectation value of energy from (\ref{energyTLSst}) is:

\begin{align}
	\begin{split}
\langle \hat{H} \rangle_A &=-\frac{\hbar\omega_2}{2}S(r_h)\tanh\Big( \frac{\hbar\omega_2}{2k_BT_h}\Big),\hspace{3mm}  \langle \hat{H} \rangle_B = -\frac{\hbar\omega_1}{2}S(r_h)\tanh \Big(\frac{\hbar\omega_1}{2k_BT_h}\Big),\\
\langle \hat{H} \rangle_C& = -\frac{\hbar\omega_1}{2}\tanh \Big(\frac{\hbar\omega_1}{2k_BT_c}\Big), \hspace{1.3cm} \langle \hat{H} \rangle_D = -\frac{\hbar\omega_2}{2}\tanh\Big( \frac{\hbar\omega_2}{2k_BT_c}\Big).
	\end{split}
\end{align}

The expectation value of work is:


\begin{align}
\begin{split}
\label{workAB}
W_{AB}&=k_BT_h\log\Bigg[\frac{\cosh\big(\frac{\hbar\omega_1}{2k_BT_h}\big)}{\cosh\big(\frac{\hbar\omega_2}{2k_BT_h}\big)}\Bigg]+\frac{1}{2}k_BT_h\log\Bigg[\frac{1+(1-S^2_{r_h})\sinh^2\big(\frac{\hbar\omega_2}{2k_BT_h}\big)}{1+(1-S^2_{r_h})\sinh^2\big(\frac{\hbar\omega_1}{2k_BT_h}\big)}\Bigg]+\\
&+G(\omega_2)\tanh\Big(\frac{\hbar\omega_2}{2k_BT_h}\Big)- G(\omega_1)\tanh\Big(\frac{\hbar\omega_1}{2k_BT_h}\Big),
\end{split}
\end{align}

where
\begin{align}
G(\omega_i)=S_{r_h} \frac{k_BT_h}{\hbar \omega_i}\log\Bigg[\frac{1+S_{r_h}+e^{-\frac{\hbar\omega_i}{k_BT_h}}(1-S_{r_h})}{1+S_{r_h}+e^{\frac{\hbar
\omega_i}{k_BT_h}}(1-S_{r_h})} \Bigg]
\end{align}

Following the same procedure as before we compute the heat and work in the third stroke
\begin{align}
Q_{CD}=T_c\Big[S(D)-S(C)\Big]=k_BT_c\log\Bigg[\frac{\cosh\Big(\frac{\hbar\omega_2}{2k_BT_c}\Big)}{\cosh\Big(\frac{\hbar\omega_1}{2k_BT_c}\Big)}\Bigg]+\frac{\hbar\omega_1}{2}\tanh\Big(\frac{\hbar\omega_1}{2k_BT_c}\Big)-\frac{\hbar\omega_2}{2}\tanh\Big(\frac{\hbar\omega_2}{2k_BT_c}\Big),
\end{align}


and
\begin{align}
	\label{workCD}
W_{CD}=k_BT_c\log\Bigg[\frac{\cosh\Big(\frac{\hbar\omega_2}{2k_BT_c}\Big)}{\cosh\Big(\frac{\hbar\omega_1}{2k_BT_c}\Big)}\Bigg].
\end{align}
In the quantum isochoric process, no work is done, thus the absorbed or released heat is:
\begin{align}
	Q_{BC}=U(C)-U(B)=\frac{\hbar\omega_1}{2}\Big[\sech(2r_h)\tanh\Big(\frac{\hbar\omega_1}{2k_BT_h}\Big)-\tanh\Big(\frac{\hbar\omega_1}{2k_BT_c}\Big)\Big],
\end{align}

\begin{align}
	Q_{DA}=U(A)-U(D)=\frac{\hbar\omega_2}{2}\Big[\tanh\Big(\frac{\hbar
	\omega_2}{2k_BT_c}\Big)-\sech(2r_h)\tanh\Big(\frac{\hbar\omega_2}{2k_BT_h}\Big)\Big].
\end{align}

The total work is computed from (\ref*{workAB}) and (\ref*{workCD}), $W=	W_{AB}+W_{CD}$, thus:

\begin{align}
\begin{split}
W=&k_BT_h\log\Bigg[\frac{\cosh\Big(\frac{\hbar\omega_1}{2k_BT_h}\Big)}{\cosh\Big(\frac{\hbar\omega_2}{2k_BT_h}\Big)}\Bigg]+k_BT_c\log\Bigg[\frac{\cosh\Big(\frac{\hbar\omega_2}{2k_BT_c}\Big)}{\cosh\Big(\frac{\omega_1}{2k_BT_c}\Big)}\Bigg]+\frac{k_BT_h}{2}\log\Bigg[\frac{1+(1-S^2_{r_h})\sinh^2\Big(\frac{\hbar\omega_2}{2k_BT_h}\Big)}{1+(1-S^2_{r_h})\sinh^2\Big(\frac{\hbar\omega_1}{2k_BT_h}\Big)}\Bigg]+\\
&+G(\omega_2)\tanh\Big(\frac{\hbar\omega_2}{2k_BT_h}\Big)- G(\omega_1)\tanh\Big(\frac{\hbar\omega_1}{2k_BT_h}\Big).
\end{split}
\end{align}

In Figs. 1 and 2 we plot the produced work $W/k_BT_c$ of the QSHE as a function of the ratio of frequencies $\omega_2/\omega_1$ and temperatures $T_h/T_C$ respectively, with different values of the squeezing parameter $r_h$. We notice that the squeezing parameter $r_h$ plays a pivotal role in the work production due to the fact that the work is maximum even if the ratio of hot and cold temperature is small, meaning less energetic cost.

 \begin{figure}[H]
\label{figwork}
\begin{center}
{{\includegraphics[scale =0.5]{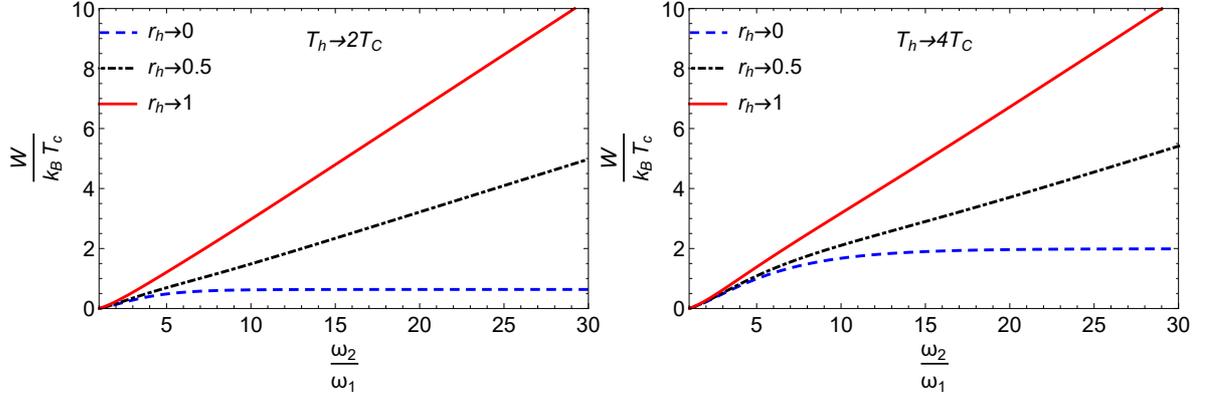}}}
\end{center}
\caption{\small Total work output $W/k_BT_c$ as a function of the frequency modulation $\omega_2/\omega_1$, for different values of the squeezed parameter $r_h$.}
\end{figure}

\begin{figure}[H]
\label{figwork}
\begin{center}
{{\includegraphics[scale =0.5]{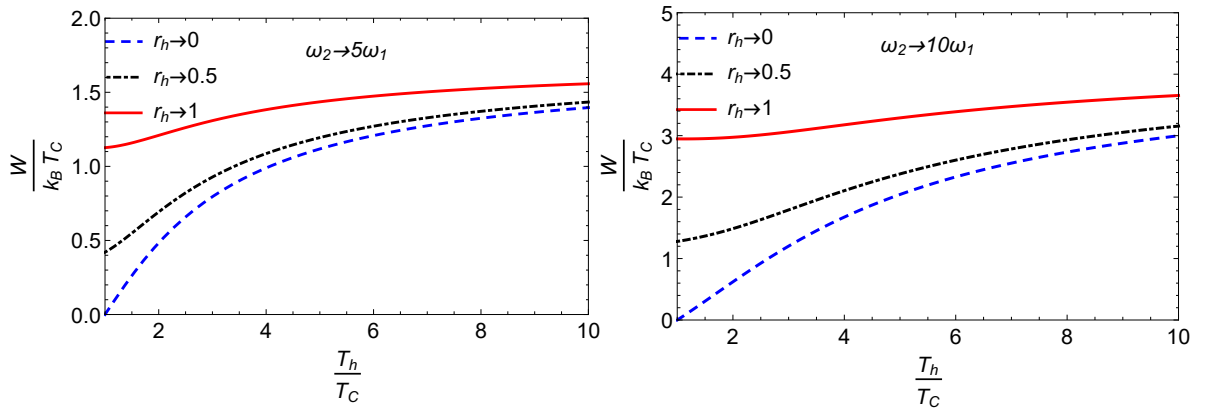}}}
\end{center}
\caption{\small Total work output $W/k_BT_c$ as a function of the temperature $T_h/T_c$, for different values of the squeezed parameter $r_h$.}
\end{figure}

The system absords heat only during the first isothermal stroke and the final isochoric stroke, thus, the total heat absorbed during a complete cycle
\begin{align}
	Q_{H}=Q_{AB}+Q_{DA}.
\end{align}

The average heat $\langle \hat{Q} \rangle_{H}\geq 0$ absorbed
from the hot squeezed reservoir and the average heat $\langle \hat{Q} \rangle_{C}\leq 0$ absorbed from the cold thermal reservoir. The mean work $\langle \hat{W} \rangle_{ab}\geq 0$ expresses the consumed work by the system and the $\langle \hat{W} \rangle_{cd}\leq 0$ means that the system produces work.

The efficiency of the QHE is the proportion of the total work
per cycle with respect to the heat absorbed from the hot thermal reservoir:

\begin{align}
	\label{EFFICIENCY STIRLING TLS}
\eta_S &= \frac{ \langle \hat{W} \rangle_{total}}{\langle \hat{Q} \rangle_{H}}=1+\frac{Q_{BC}+Q_{CD}}{	Q_{AB}+Q_{DA}}.
\end{align}

In Fig. 3, we plot the efficiency $\eta_{S}$ of our cycle as a function of the squeezing parameter $r_h$ and compare it with Carnot and Curzon-Ahlborn efficiency \cite{Curzon}. We notice that the squeezing parameter $r_h$ increases the efficiency, especially in the small ratio of temperatures, and the efficiency tends to Curzon-Ahlborn efficiency $\eta_{CA}$, which is the  efficiency at maximum work \cite{Broeck}. On the other hand, when the ratio of temperatures $T_h/T_C$ increases the squeezing parameter can not improve the performance of QSHE so as to surpass the Carnot efficiency $\eta_{C}$.

\begin{figure}[H]
\begin{center}
	{{\includegraphics[scale =0.53]{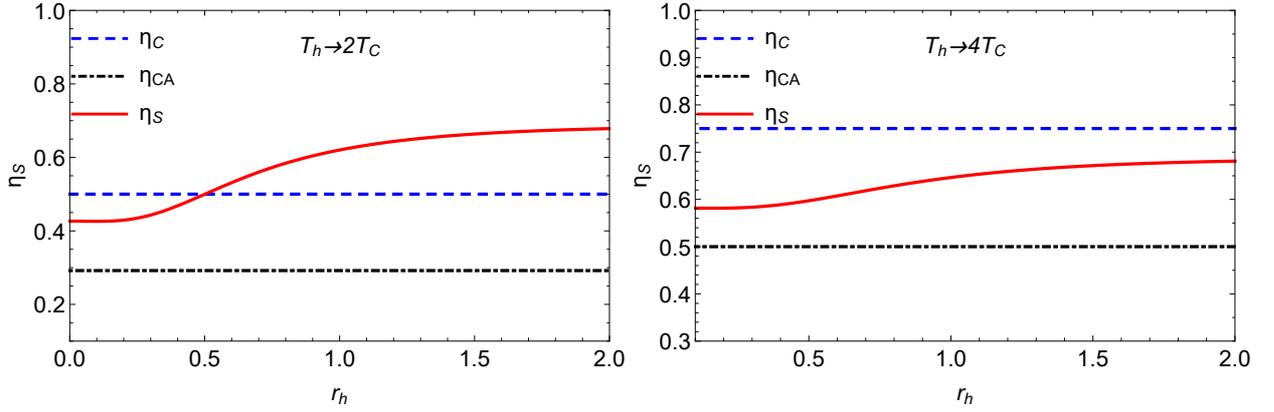}}}%
	\end{center}
	\caption{\small Efficiency of the heat engine as a function of the squeezing parameter $r_h$ in comparison to Carnot $\eta_C$ and Curzon Ahlborn $\eta_{CA}$ efficiency. Others parameters are $\omega_2=5\omega_1$.}
\end{figure}

\begin{figure}[H]
\begin{center}
	{{\includegraphics[scale =0.53]{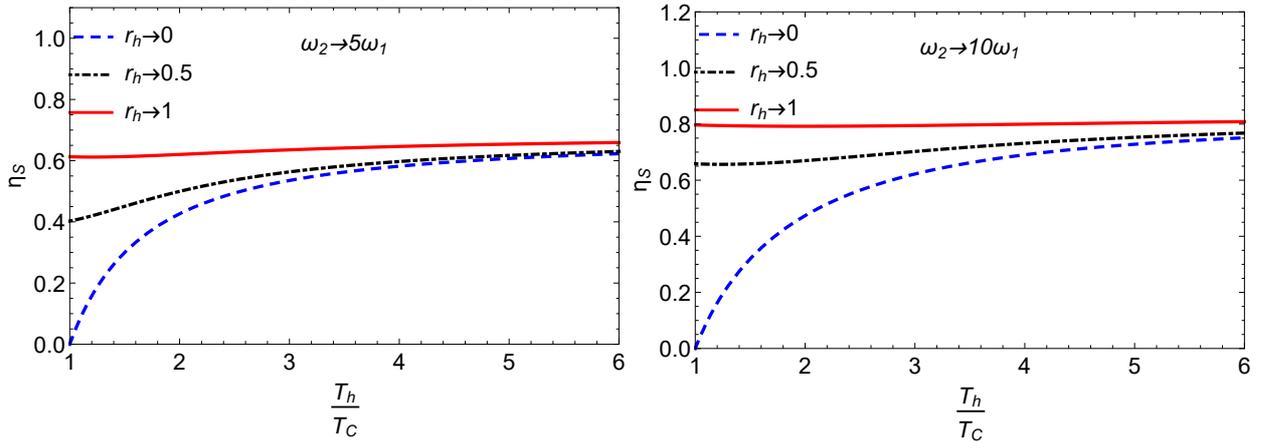}}}%
	\end{center}
	\caption{\small Efficiency of the heat engine as a function of the temperatures $T_c/T_h$ for different ratio of frequencies.}
\end{figure}

As indicated in Fig. 5 the efficiency of the QSHE increases with the increase of the ratio $\omega_2/\omega_1$ and the $r_h$.

\begin{figure}[H]
\begin{center}
	{{\includegraphics[scale =0.9]{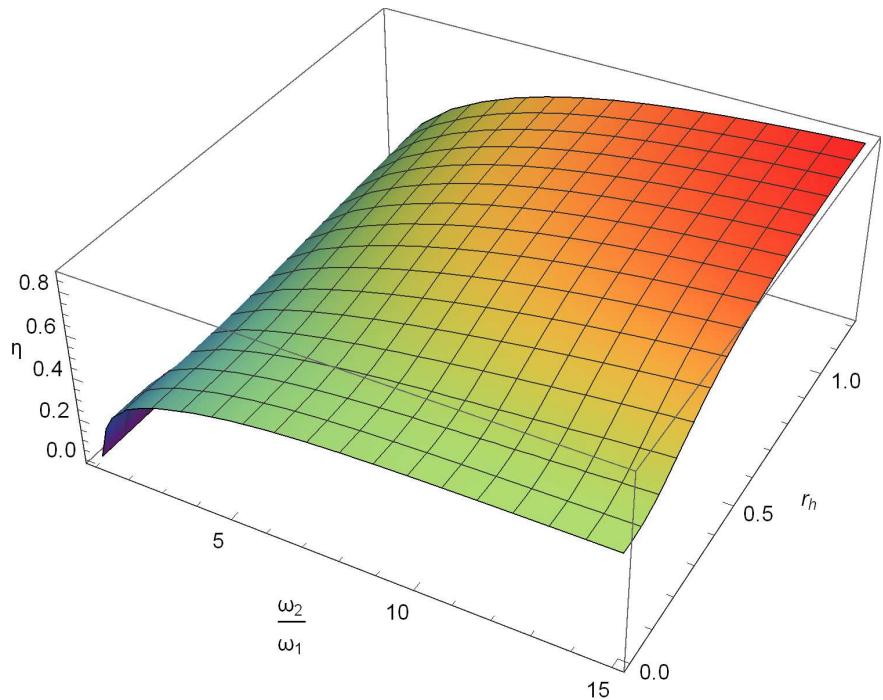}}}%
	\end{center}
	\caption{\small Efficiency of the heat engine as a function of $\omega_2/\omega_1$ and the squeezing parameter $r_h$ for $T_h=2T_C$.}
\end{figure}

We highlight that the efficiency of the QSHE with squeezed thermal reservoir surpasses the standard Carnot limit when the ratio of hot and cold bath is small. This interesting result is caused when the squeezing parameter $r_h \rightarrow 1$, and its mechanisms can be explained as follows. At the end of the quantum isochoric process, the asymptotic state of the two level system is given from (\ref{asymptoticTLS}). When $r_h \rightarrow 0$ the asymptotic state of the two level system is given from,

\begin{eqnarray}
	\label{asymptoticTLSr0}
	\rho = \frac{1}{(2n+1)}
	\bordermatrix{&              &              \cr
		& n
		& 0 \cr
		& 0
		& n+1\cr},
\end{eqnarray}
where $n$ is given from (\ref{n}). On the contrary, when $r_h > 0$, the asymptotic state of the two level system is given again from (\ref{asymptoticTLSr0}) but with an effective temperature that contains the squeezing parameter $r_h$ and the temperature T,


\begin{align}
T^{eff}=\frac{\hbar\omega}{2k_B\arctanh\Big[\frac{\tanh\big(\frac{\hbar\omega}{2k_BT}\big)}{\cosh (2r_h)}\Big]},
\end{align}

and $T^{eff}$ is always higher than T due to the squeezed parameters $r_h$ \cite{HuangEffects}.

Let us now, proceed to examine the {\em high temperature limit: $\frac{\omega}{T_h} \rightarrow 0$}. The total produced work in the first order approximation is:

\begin{align}
W_{ht}=\frac{\hbar^2(\omega_2^2-\omega_1^2)}{8k_B}\Big[S_{r_h}(S_{r_h}-2)\frac{1}{T_h}+\frac{1}{T_C}\Big].
\end{align}

 We use the second derivative test with respect to $\omega_2$ \cite{Abah} in order to find the maximum value of the produced work. In this case the produced work does not have a local maximum value.
 
Next, we compute the efficiency in the high temperature regime in the second order approximation:

\begin{align}
\eta_{mw}^{ht}=1+\frac{2\frac{\omega_1^2}{\omega_2^2}S_{r_h}(1-\eta_C)-1)-(1-\frac{\omega_1^2}{\omega_2^2})}{2-(1-\eta_C)\hbar S_{r_h}\Big[2-\hbar S_{r_h}(1-\eta_C)(1-\frac{\omega_1^2}{\omega_2^2})\Big]}.
\end{align}

Let us now, proceed to examine the {\em  low temperature limit: $\frac{\omega}{T_C} \rightarrow \infty$}. The total produced work is:

\begin{align}
W_{lt}=\frac{1}{2}\hbar(\omega_2-\omega_1)+\frac{\hbar^2(\omega_2^2-\omega_1^2)}{8k_BT_h}\Big[S_{r_h}(S_{r_h}-2)\Big],
\end{align}
where the first and the second term is the first and second order approximation respectively. The maximum work is when

\begin{align}
\label{omegaTLS}
\omega_2=\frac{2k_BT_h}{\hbar}\frac{1}{4-3S_{r_h}^{2}}.
\end{align}

The efficiency in the low temperature regime is

\begin{align}
\label{otto}
\eta_{mw}^{lt}=1-\frac{\omega_1}{\omega_2},
\end{align}
which coincides with the efficiency of quantum Otto heat engine \cite{Assis}, \cite{Assis1} and is independent of the squeezing parameter $r_h$ and depends on the ratio of the frequencies only. The efficiency at maximum work is taken from (\ref{omegaTLS}) and (\ref{otto}).

The second order approximation of the efficiency, which depends on squeezing parameter $r_h$, is:

\begin{align}
\eta_{mw}^{lt}=1-\frac{\omega_1^2}{\omega_2^2}\frac{\frac{4k_BT_h}{\omega_1}-2\hbar S_{r_h}}{\frac{4k_BT_h}{\omega_2}-\hbar S_{r_h}\Big[2-S_{r_h}(1-\frac{\omega_1^2}{\omega_2^2})\Big]}
\end{align}

\section{Quantum Stirling heat engine - Harmonic oscillator}
\label{Quantum Stirling heat engine - HO}

In this section we examine the harmonic oscillator as the working medium of the QSHE. At first, we refer the thermal squeezed state of the harmonic oscillator. For a harmonic oscillator of mass $m$ and frequency $\omega$, the squeezed thermal state results from the application of the squeezing operator to the thermal equilibrium state. The density matrix is as follows:

     \begin{align}
         \rho_{sq}=\hat{S}(\xi)\rho_G\hat{S}^\dagger(\xi)=\hat{S}(\xi)\frac{e^{-\beta \hat{H}}}{Z(\beta)}\hat{S}^\dagger(\xi)
     \end{align}
where $\hat{\rho}_G$ is the Gibbs state of the harmonic oscillator, and $\hat{S}(\xi)$ is the squeezed operator,
\begin{equation}
    \hat{S}(\xi)=e^{\frac{1}{2}(\xi^{*\alpha^{2}}-\xi\alpha^{\dagger^2})},
\end{equation}

and
\begin{align*}
          \xi=re^{i\theta}, 
     \end{align*}
     
is a complex number with $r\geq 0$  called the squeezing parameter with $\theta \in [0,2\pi]$.\\
 The internal energy of the harmonic oscillator in a squeezed thermal state is \cite{BEI}

\begin{align}
\langle \hat{H} \rangle &=\hbar\omega\Big(n+\frac{1}{2}\Big)\cosh (2r_h),
\end{align}
where
\begin{align}
n &=\frac{1}{e^{\frac{\hbar \omega}{k_BT}}-1}.
\end{align}

In an analogous way, as in the first part of this work, we construct the four strokes of the QSHE with the harmonic oscillator as the working medium \cite{Yair Rezek and Ronnie Kosloff}, and then we compute the expectation value of energy of each process, thus:

\begin{align}
	\begin{split}
\langle \hat{H} \rangle_A &=\frac{\hbar\omega_2}{2S(r_h)}\coth\Big( \frac{\hbar\omega_2}{2k_BT_h}\Big),\hspace{3mm}  \langle \hat{H} \rangle_B = \frac{\hbar\omega_1}{2S(r_h)}\coth\Big( \frac{\hbar\omega_1}{2k_BT_h}\Big),\\
\langle \hat{H} \rangle_C& = \frac{\hbar\omega_1}{2}\coth\Big( \frac{\hbar\omega_1}{2k_BT_c}\Big), \hspace{0.85cm} \langle \hat{H} \rangle_D = \frac{\omega_2}{2}\coth\Big( \frac{\hbar\omega_2}{2k_BT_c}\Big).
	\end{split}
\end{align}

Consequently, the heat in the isothermal process $AB$ is:


\begin{align}
\label{QABSHO}
\begin{split}
    Q_{AB}&=k_BT_h\sum_{i=1}^{2}(-1)^{i+1}\Big\{F_{r_h}(\omega_i)\log [F_{r_h}(\omega_i)]-(F_{r_h}(\omega_i)-1)\log [F_{r_h}(\omega_i)-1]\Big\},
\end{split}
\end{align}

where

\begin{align}
F_{r_h}(\omega)=\frac{1}{2}\Bigg[1+S^{-1}_{r_h}\coth\Big(\frac{\hbar\omega}{2k_BT_h}\Big)\Bigg].
\end{align}

The expectation value of work in this process is obtained from the first law of thermodynamics:

\begin{align}
\begin{split}
\label{workABHO}
W_{AB}&=k_BT_h\sum_{i=1}^{2}(-1)^{i+1}\Big\{F_{r_h}(\omega_i)\log [F_{r_h}(\omega_i)]-(F_{r_h}(\omega_i)-1)\log [F_{r_h}(\omega_i)-1]\Big\}+\\
&+\frac{\hbar\omega_2}{2S(r_h)}\coth\Big( \frac{\hbar\omega_2}{2k_BT_h}\Big)-\frac{\hbar\omega_1}{2S(r_h)}\coth\Big( \frac{\hbar\omega_1}{2k_BT_h}\Big).
\end{split}
\end{align}

In the quantum isochoric processes, no work is done, thus the absorbed or released heat is:
\begin{align}
	Q_{BC}=U(C)-U(B)=\frac{\hbar\omega_1}{2}\Bigg[\coth\Big(\frac{\hbar\omega_1}{2k_BT_c}\Big)-\frac{1}{S(r_h)}\coth\Big(\frac{\hbar\omega_1}{2k_BT_h}\Big)\Bigg]
\end{align}

\begin{align}
	Q_{DA}=U(A)-U(D)=\frac{\hbar\omega_2}{2}\Bigg[\frac{1}{S(r_h)}\coth\Big(\frac{\hbar\omega_2}{2k_BT_h}\Big)-\coth\Big(\frac{\hbar\omega_2}{2k_BT_c}\Big)\Bigg].
\end{align}

Following the same procedure as before we compute the heat and work in the third stroke
\begin{align}
Q_{CD}=T_c\Big[S(D)-S(C)\Big]=k_BT_c\log\Bigg[\frac{\sinh\Big(\frac{\hbar\omega_1}{2k_BT_c}\Big)}{\sinh\Big(\frac{\hbar\omega_2}{2k_BT_c}\Big)}\Bigg]+\frac{\hbar\omega_2}{2}\coth\Big(\frac{\hbar\omega_2}{2k_BT_c}\Big)-\frac{\hbar\omega_1}{2}\coth\Big(\frac{\hbar\omega_1}{2k_BT_c}\Big).
\end{align}

Consequently, the total produced work is:

\begin{align}
\begin{split}
\label{workHO}
W_{total}&=k_BT_h\sum_{i=1}^{2}(-1)^{i+1}\Big\{F_{r_h}(\omega_i)\log [F_{r_h}(\omega_i)]-(F_{r_h}(\omega_i)-1)\log [F_{r_h}(\omega_i)-1]\Big\}+\\
+&k_BT_c\log\Bigg[\frac{\sinh\Big(\frac{\hbar\omega_1}{2k_BT_c}\Big)}{\sinh\Big(\frac{\hbar\omega_2}{2k_BT_c}\Big)}\Bigg].
\end{split}
\end{align}

In Figs. 6 and 7, we show that the total produced work increases monotonically from zero as a function of the frequency ration respectively with the increase of the squeezing parameter $r_{h}$.

\begin{figure}[H]
\label{figworkHO}
\begin{center}
{{\includegraphics[scale =0.5]{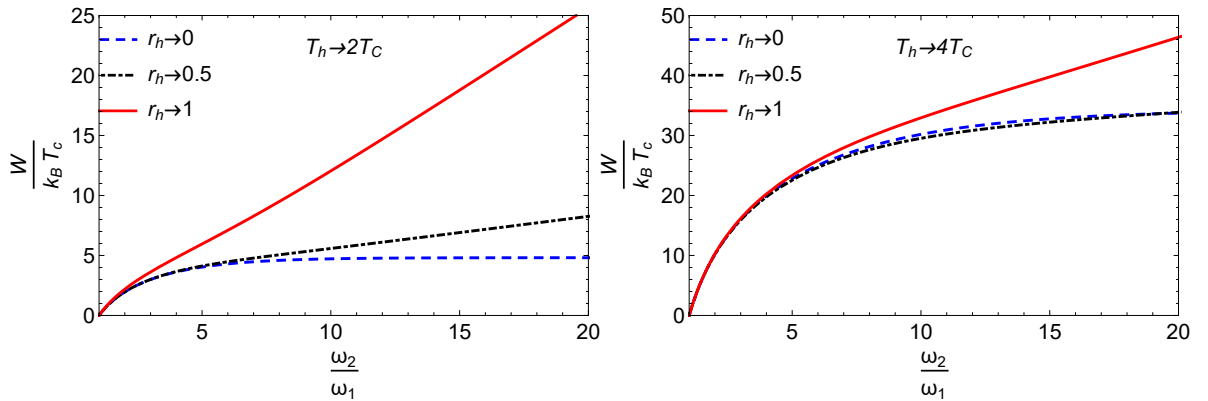}}}
\end{center}
\caption{\small Total work output $W/k_BT_c$ as a function of the frequency modulation, $\omega_2/\omega_1$, for different values of the squeezed parameter $r_h$ and temperature ratio.}
\end{figure}

\begin{figure}[H]
\label{figwork}
\begin{center}
	{{\includegraphics[scale =0.53]{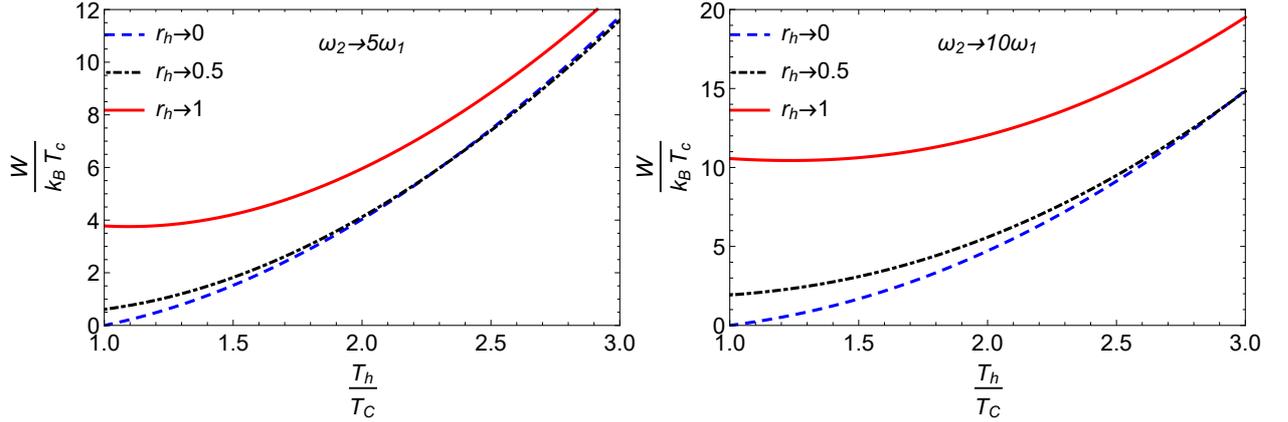}}}%
	\end{center}
	\caption{\small Total work output, $W/k_BT_c$ as a function of temperatures ratio, $T_h/T_C$,
 for different values of the squeezed parameter $r_h$ and frequency ratio.}
\end{figure}

Furthermore, we compute the efficiency of the QSHE

\begin{align}
	\label{EFFICIENCY STIRLING HO}
\eta &= \frac{ \langle \hat{W} \rangle_{total}}{\langle \hat{Q} \rangle_{H}}=1+\frac{Q_{BC}+Q_{CD}}{Q_{AB}+Q_{DA}}.
\end{align}
Due to the fact that the analytical expression of the efficiency of the QSHE is complicated, we plot the numerical result in Figs. 8 and 9. In Fig. 8 we plot the efficiency as a function to squeezing parameter $r_h$ and compare it with the  Curzon Ahlborn $\eta_{CA}$ efficiency and the efficiency of the classical Stirling heat engine, which is the same with the Carnot efficiency $\eta_C$. We show that the increase of $r_h$ can improve the efficiency of the QSHE so as to approach and surpass the Carnot limit.

\begin{figure}[H]
\label{figwork}
\begin{center}
	{{\includegraphics[scale =0.53]{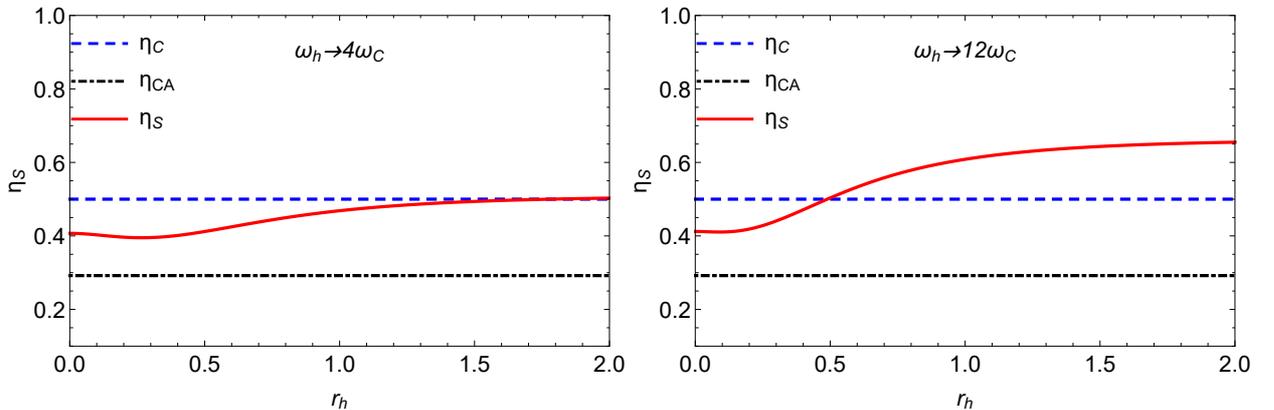}}}%
	\end{center}
	\caption{\small Efficiency $\eta$  as a function of the squeezed parameter $r_h$ for different frequencies of the oscillator $\omega_1=k_BT_c/\hbar$ in comparison to Carnot $\eta_C$ and Curzon Ahlborn $\eta_{CA}$ efficiency. Other parameters are $T_h=2T_c$}
\end{figure}
In Fig. 9 we plot the efficiency of the QSHE as a function of the squeezing parameter $r_h$ and the ratio of frequencies $\omega_2/\omega_1$. We show that the efficiency increases with the increase of $\omega_1=k_BT_c/\hbar$ and also with $r_h$.
\begin{figure}[H]
\begin{center}
	{{\includegraphics[scale =0.9]{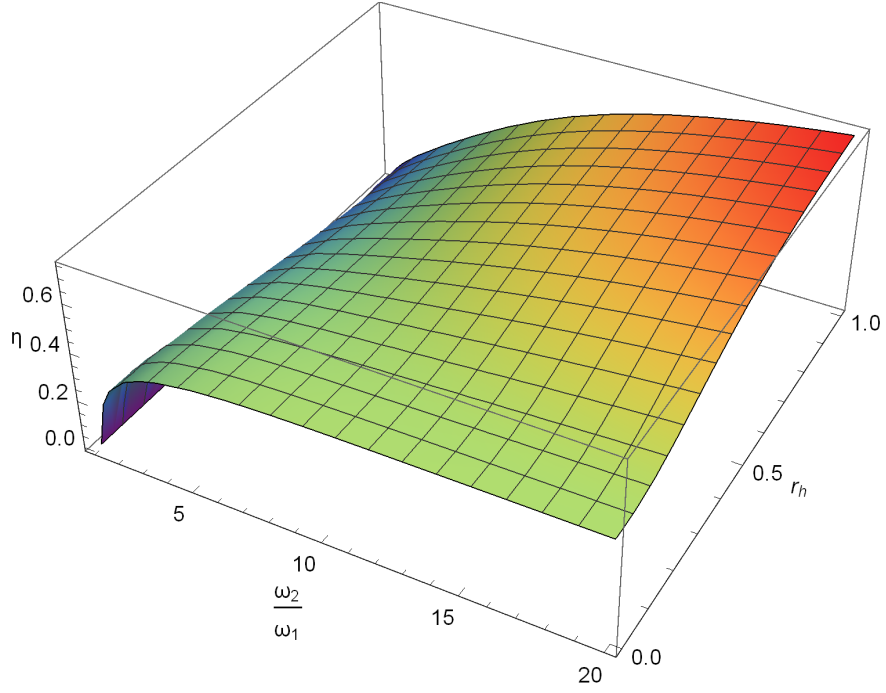}}}%
	\end{center}
	\caption{\small Efficiency of QSHE with respect to $\omega_2/\omega_1$ and the squeezing parameter $r_h$. The oscillator 's frequency is $\omega_1=k_BT_c/\hbar$.}
\end{figure}

Let us now, proceed to examine the {\em  high temperature limit: $\frac{\omega}{T_h} \rightarrow 0$}. The total produced work is:

\begin{align}
W_{ht}=k_B(T_h-T_C)\log\Big(\frac{\omega_2}{\omega_1}\Big)+\frac{\hbar}{12k_BT_hS_{r_h}}(\omega_2^2-\omega_1^2)+\frac{\hbar^2}{24}(\omega_1^2-\omega_2^2)\Big(\frac{S_{r_h}}{k_BT_hS_{2r_h}}+\frac{1}{k_BT_C}\Big)
\end{align}
where the first term is the first order approximation and the rest terms are derived from the second-order approximation.
Following the same procedure as in the first part of this work, we find that the maximum work is when

\begin{align}
\omega_2=\frac{2\sqrt{3k_BT_C\eta_C}}{\hbar \sqrt{1+S_{2r_h}^{-1}S_{r_h}^2(1-\eta_C)}}.
\end{align}
Consequently, the efficiency at maximum work in the high temperature regime is: 

\begin{align}
\eta_{mw}^{ht}=\frac{\eta_C}{\eta_C-1+S_{r_h}^{-1}+\log\Big(\frac{\omega_2}{\omega_1}\Big)}\log\Big(\frac{\omega_2}{\omega_1}\Big).
\end{align}
Let us now, proceed to examine the {\em low temperature limit: $\frac{\omega}{T_C} \rightarrow \infty$}. The total produced work is:

\begin{align}
W_{lt}=k_BT_h\log\Big(\frac{\omega_2}{\omega_1}\Big)+\frac{1}{2}\hbar(\omega_1-\omega_2)+\frac{\hbar}{12}(\omega_1^2-\omega_2^2)\Big(\frac{1}{S_{r_h}}-\frac{S_{r_h}}{2S_{2r_h}}\frac{\hbar}{k_BT_h}\Big)
\end{align}
where the first two terms are the first order approximation.

The maximum work is when

\begin{align}
\omega_2=\frac{1+S_{2r_h}}{2}\frac{k_BT_h}{\hbar}\Big[-3+\sqrt{3(11-4S_{r_h}^2)}\Big]
\end{align}

Consequently, the efficiency at maximum work in the low temperature regime is:

\begin{align}
\eta_{mw}^{lt}=1+\frac{\hbar\omega_1-2k_BT_hS^{-1}_{r_{h}}}{-\hbar\omega_2+2k_BT_h[S^{-1}_{r_{h}}+\log\big(\frac{\omega_2}{\omega_1}\big)]}
\end{align}

\section{Conclusions}
\label{Conclusions}
In this paper, we investigated a quantum Stirling heat engine (QSHE) with a two level system and a harmonic oscillator as the working medium that contacts with a squeezed thermal reservoir and a cold reservoir. We obtained closed-form expressions for the produced work and efficiency which depends strongly on the squeezing parameter $r_h$. Then we proved that the effect of squeezing heats the working medium to a higher effective temperature leading to a better overall performance. In particular, the efficiency increases with the degree of squeezing surpassing the standard Carnot limit, when the ratio of temperatures of hot and cold reservoir is small, which means that the engine has less energetic cost. Additionally, we derived the analytic expression for the efficiency at maximum work and the maximum produced work in the high and low temperature regime. We found that the efficiency in the low temperature regime coincides with the efficiency of quantum Otto heat engine and is independent of the squeezing parameter $r_h$, depending only on the ratio of frequencies. We noticed that in the first order approximation the efficiency at maximum work and the maximum produced work is independent of the squeezing parameter $r_h$. On the contrary, the aforementioned result in the second order approximation depends on the squeezing parameter $r_h$. This means that at extreme temperatures the squeezing parameter $r_h$ does not affect the performance of the QSHE. Finally, the performance of the QSHE depends on the nature of the working medium.


\end{document}